\documentclass[english]{article}
\usepackage[T1]{fontenc}
\usepackage[latin9]{inputenc}
\usepackage{geometry}
\usepackage{babel}
\usepackage{amsmath}
\usepackage{amssymb}
\usepackage{esint}
\usepackage[unicode=true]
 {hyperref}

\makeatletter

\newcommand{\mathcircumflex}[0]{\mbox{\^{}}}

\@ifundefined{definecolor}
 {\usepackage{color}}{}
\@ifundefined{definecolor}{\usepackage{color}}{}

\usepackage{babel}
\usepackage{nicefrac}
\usepackage{sagetex}

\makeatother

\begin{document}

\title{Integer formula encoding\\
 $\textsf{SageTeX}$ package}

\author{Edinah K. Gnang}
\maketitle
\begin{abstract}
The following $\textsf{SageTeX}$ document accompanies the papers
\cite{GZ,GRS}, available from Gnang's websites. Please report bugs
to gnang at cs dot rutgers dot edu. The most current version of the
$\textsf{SageTeX}$ document are available from \href{http://www.cs.rutgers.edu/~gnang}{Gnang's website} 
\end{abstract}

\section{Introduction}

As quoted by Weber (1893), Leopold Kronecker is known to have said:
\textquotedbl{}\emph{God made natural numbers; all else is the work
of man}\textquotedbl{}. The proposed packages adresses aspects of
combinatorial aspects of integer encodings and can be paraphrased
as a slight modification of Kronecker's quote: \emph{God created the
integral unit} ``$1$''; \emph{all else is the result of computation}.
The topic of integer encoding schemes, is one which generates interest
both from the amateur and professional mathematician alike, since
:
\begin{itemize}
\item barriers to entry to the subject are virtually non existent, in light
of the fact that the main ideas can easily be conveyed to elementary
school students.
\item it's topics have ramifications and connections with other topics in
mathematics such as algebra, cobinatorics and number theory. 
\item most importantly, the topic offers a treasure trove of fascinating
easy to state open questions. 
\end{itemize}
A number \emph{circuit} encoding $\varPhi$ is a finite directed acyclic
graph constructed as follows. Nodes of in-degree zero are labeled
by either of the constants $1$ or $\left(-1\right)$. All other nodes
of the graph have in-degree two and are labeled either ($+$), ($\times$)
or ($\mathcircumflex$). The two edges going into a gate labeled by
($\mathcircumflex$) are labelled by \emph{left} and \emph{right},
in order to distinguish the base (left input) from the exponent (right
input). The nodes of out-degree zero correspond to output gates of
the circuit. 

The \emph{size} of $\varPhi$ is the number of nodes in $\varPhi$.
The \emph{depth }of $\varPhi$ is the length of the longest path in
$\varPhi$. A number \emph{formula }encoding is a special circuit
with the additional restriction that every node has out-degree at
most one. Given an monotonically increasing function $\mathfrak{s}$
\[
\mathfrak{s}\::\:\mathbb{N}\rightarrow\mathbb{N}
\]
we seek to determine the number of formula encodings for some integer
$n$ of size at most $\mathfrak{s}\left(n\right)$. In many cases
the analysis is considerably simplified by considering \emph{monotone}
formula encodings, namely formula encodings further restricted to
have all in-degree zero nodes labeled with the constant $1$. It is
rather natural to consider formulas for which $1$ is never an input
to a multiplication or an exponentiation gate. It was shown in \cite{GRS}
there exists constants $c>0$ and $\rho>4$ such that some real number
number $\rho>4$ such that the number of formula encodings of $n$
is asymptotically equal to 
\begin{equation}
c\,\frac{\rho^{n}}{\left(\sqrt{n}\right)^{3}}
\end{equation}
In the more general setting where the label $\left(-1\right)$ is
allowed for in-zero nodes of the graphs the asymptotics for the number
formula encodings for an integer $n$ of size not exceeding $\left(2n-1\right)$
as $n$ tends to infinity is still unknown. The content of this paper
is the following. In section 2 we provide a general overview of the
computational model and our basic assumptions. The rest of the paper
provides an annotated implementations of the various procedures for
manipulating formulas encodings. A seperate sage file which isolates
the procedures acompanies the paper and can be used for experimental
set up with our proposed package.

\section{Basic overview of the integer formula encoding model}

Let $\mathcal{F}$ denote the set of formula encodings constructed
by combining finitely many fan-in two addition ($+$), multiplication
($\times$) and exponentiation ($\mathcircumflex$) gates with restricted
to either constants $1$ or $-1$. For the sake of completeness we
pin down our computational model by describing formula transformation
rules which prescribe equivalences among distinct elements of $\mathcal{F}$.
Let $f$, $g$, and $h$ denote arbitrary elements of $\mathcal{F}$.
The equivalence between distinct elements of $\mathcal{F}$ is prescribed
by the following transformation rules
\begin{enumerate}
\item Commutativity 
\begin{equation}
\begin{array}{c}
f+g\leftrightarrows g+f\\
f\times g\leftrightarrows g\times f
\end{array}
\end{equation}

\item Associativity 
\begin{equation}
\begin{array}{c}
\left(f+g\right)+h\leftrightarrows f+\left(g+h\right)\\
\left(f\times g\right)\times h\leftrightarrows f\times\left(g\times h\right)
\end{array}
\end{equation}

\item Unit element 
\begin{equation}
\begin{array}{ccc}
f\times1 & \leftrightarrows & f\\
f\mathcircumflex1 & \leftrightarrows & f\\
1\mathcircumflex f & \leftrightarrows & 1\\
f+\left(1+\left(-1\right)\right) & \leftrightarrows & f\\
f\times\left(1+\left(-1\right)\right) & \leftrightarrows & \left(1+\left(-1\right)\right)\\
f\mathcircumflex\left(1+\left(-1\right)\right) & \leftrightarrows & 1
\end{array}
\end{equation}

\item Distributivity
\begin{equation}
\begin{array}{c}
f\times\left(g+h\right)\leftrightarrows f\times g+f\times h\\
f\mathcircumflex\left(g+h\right)\leftrightarrows f\mathcircumflex g\times f\mathcircumflex h\\
\left(f\times g\right)\mathcircumflex h\leftrightarrows f\mathcircumflex h\times g\mathcircumflex h
\end{array}.
\end{equation}

\end{enumerate}
Finally an important rule is that a formula is considered invalid
if admits as a subformula any formula equivalent to $\left(1+\left(-1\right)\right)\mathcircumflex\left(-1\right)$
via the transformation rules prescribed above. Throughout the discussion,
the efficiency of formula encodings will be a recurring theme and
thus we (often implicitly) exclude from $\mathcal{F}$ formulas which
admit sub-formulas of the form 
\[
1\times f,\quad f\times1,\quad f\mathcircumflex1,\quad1\mathcircumflex f.
\]
We remark as is well known that any formulas from the set $\mathcal{F}$
can be uniquely encoded as strings from the the alphabet
\begin{equation}
\mathfrak{A}\;:=\left\{ 1,\:-1,\:+,\:\times,\:\mathcircumflex\right\} ,
\end{equation}
using either the prefix or the postfix/polish notation.

Let $C_{S_{1}}^{S_{0}}\left(n\right)$ denotes the number of formulas
encoding in $\mathcal{F}$ which evaluated to $n$ and of size not
exceeding $\left(2n-1\right)$ constructed using gates from the set
$S_{1}$ and rooted at any of the gates in the set $S_{0}$ where
$S_{0}\subseteq S_{1}\subseteq\left\{ +,\times,\mathcircumflex\right\} $.
As pointed in \cite{GZ,GRS}, the non linear recurrence relations
which determines the counts for the number of formulas encodings of
$n$ and incidentally the number of vertices of the equivalence class
graph associated with the integer $n$ is given by 
\begin{equation}
C_{\left\{ +,\times,\mathcircumflex\right\} }^{\left\{ +\right\} }\left(n\right)=\sum_{i}C_{\left\{ +,\times,\mathcircumflex\right\} }^{\left\{ +,\times,\mathcircumflex\right\} }\left(i\right)\: C_{\left\{ +,\times,\mathcircumflex\right\} }^{\left\{ +,\times,\mathcircumflex\right\} }\left(n-i\right)
\end{equation}
\begin{equation}
C_{\left\{ +,\times,\mathcircumflex\right\} }^{\left\{ \times\right\} }\left(n\right)=\sum_{i}C_{\left\{ +,\times,\mathcircumflex\right\} }^{\left\{ +,\times,\mathcircumflex\right\} }\left(i\right)\: C_{\left\{ +,\times,\mathcircumflex\right\} }^{\left\{ +,\times,\mathcircumflex\right\} }\left(i^{(-1)}n\right)
\end{equation}
\begin{equation}
C_{\left\{ +,\times,\mathcircumflex\right\} }^{\left\{ \mathcircumflex\right\} }\left(n\right)=\sum_{i}C_{\left\{ +,\times,\mathcircumflex\right\} }^{\left\{ +,\times,\mathcircumflex\right\} }\left(i\right)\: C_{\left\{ +,\times,\mathcircumflex\right\} }^{\left\{ +,\times,\mathcircumflex\right\} }\left(n^{i^{(-1)}}\right)
\end{equation}
and 
\begin{equation}
C_{\left\{ +,\times,\mathcircumflex\right\} }^{\left\{ +,\times,\mathcircumflex\right\} }\left(n\right)=\sum_{\mathfrak{g}\in\left\{ +,\times,\mathcircumflex\right\} }C_{\left\{ +,\times,\mathcircumflex\right\} }^{\left\{ \mathfrak{g}\right\} }\left(n\right)
\end{equation}
In order to analyze arithmetic algorithms, we introduce the graph
$G_{n}$ whose vertices are elements $\mathcal{F}$ which belong to
the equivalence class of formulas of size at most $\left(2n-1\right)$
which evaluate to some given number $n$. We shall refer to $G_{n}$
as the arithmeticahedron of $n$. Edges are placed in between any
two vertices of $G_{n}$ if either of the following conditions are
true
\begin{enumerate}
\item Each formula vertex can be obtained from the other by the use of a
single associativity transformation rules.
\item Each formula vertex can be obtained from the other by the use of a
single commutativity transformation rule.
\item Each formula can be obtained from the other by the use of one of the
distributivity transformation rules.
\end{enumerate}
Arithmetical algorithm can thus be depicted as walks on some arithmeticahedron
and incidentally the performance of algorithm can be measured in terms
of the total length of walks on some arithmeticahedron.

\section{Listing integer monotone formula encodings}

We present here the implementation details of our integer encoding
packages. The package will be crucial for setting up various experiments
which would suggest interesting conjecture and possibly proofs to
some of these conjectures. We shall think of our formulas as rooted
binary trees with leafs labeled with the integral unit ($1$) and
all other vertices labeled with either the addition ($+$), multiplication
($\times$), or exponentiation ($\mathcircumflex$) operation. It
shall be convenient to use the bracket notation to specify such trees
to sage and note that the prefix notation is easily obtain from the
bracket notation.\\
\begin{sageblock}
def T2Pre(expr):
    """
    Converts formula written in the bracket tree encoding to the 
    Prefix string encoding notation


    EXAMPLES:
    The implementation here tacitly assumes that the input 
    is a valid binary bracket formula-tree expression. The usage of the function
    is illustrated bellow.
    ::
        sage: T2Pre(['+',1,1])
        '+11'


    AUTHORS:
    - Edinah K. Gnang and Doron Zeilberger

    To Do :
    - 

    """
    s = str(expr)
    return ((((s.replace("[","")).replace("]",""))\
.replace(",","")).replace("'","")).replace(" ","")
\end{sageblock}\\
As the code for the function T2Pre suggest the binary-tree formula
is very close to the prefix notation. The usage of the function is
illustrated bellow
\begin{equation}
\mbox{T2Pre}\left(['+',1,1]\right)=\mbox{'+11'}
\end{equation}
A minor variation on the prefix notation called the postfix notation
is implemented bellow\\
\begin{sageblock}
def T2P(expr):
    """
    The function converts binary formula-tree format to the more compacts
    postfix string notation.


    EXAMPLES:
    The implementation here tacitly assumes that the input 
    is a valid binary formula-tree expression. The usage of the function
    is illustrated bellow.
    ::
        sage: T2P(['+',1,1])
        '11+'


    AUTHORS:
    - Edinah K. Gnang, Maksym Radziwill and Doron Zeilberger

    To Do :
    - 

    """
    s = str(expr)
    return ((((s.replace("[","")).replace("]",""))\
.replace(",","")).replace("'","")).replace(" ","")[::-1]
\end{sageblock}The usage of the function is illustrated bellow
\begin{equation}
\mbox{T2P}\left(['+',1,1]\right)=\mbox{'11+'}
\end{equation}
When using the Wilf Methodology \cite{NW}, we will require a random
number generator which amounts to rolling a loaded die. We implement
here the function allowing us to roll a loaded die.\\
\begin{sageblock}
def RollLD(L):
    """
    The functions constructs a loaded die according to values specified
    by the input list of positive integers. The input list also 
	specifies the desired bias for each one of the faces of the dice


    EXAMPLES:
    The tacitly assume that the input list is indeed made up of positive integers
    as no check is perform to validate that assumption 
    ::
        sage: RollLD([1, 2, 3])
        2


    AUTHORS:
    - Edinah K. Gnang, Maksym Radziwill and Doron Zeilberger

    To Do :
    - Try to implement faster version of this procedure

    """
    # Summing up all the 
    N = sum(L)
    r = randint(1,N)
    for i in range(len(L)):
        if sum(L[:i+1]) >= r:
            return i+1
	
\end{sageblock}\\
Given a list of positive integers the procedures operates in two steps.
First it samples uniformly at random a positive integer less than
the sum of all the positive integers in the input list. The last step
consist in returning the largest index of the element in the input
list such that the sum of the integers preceding that index is less
or equal to the sampled integers.

\subsection{Formulas only using additions}

We provide here a straight forward implementation of procedures for
listing formulas which only uses addition. \\
\begin{sageblock}
@cached_function
def FaT(n):
    """
    The procedure outputs the list of Formula-binary Trees
    constructed using fan-in two addition gates and having
    inputs restricted to the integral unit 1 and the resulting
    formulas each evaluate to the input integer n > 0.

    EXAMPLES:
    The procedure expects a positive integer otherwise
    it returns the empty list.
    ::
        sage: FaT(3)
        [['+', 1, ['+', 1, 1]], ['+', ['+', 1, 1], 1]]


    AUTHORS:
    - Edinah K. Gnang, Maksym Radziwill and Doron Zeilberger

    To Do :
    - 

    """
    if n==1:
        return [1]
    elif n > 1 and type(n) == Integer:
        gu = []
        for i in range(1,n):
            gu = gu + [['+', g1, g2] for g1 in FaT(i) for g2 in FaT(n-i)]
        return gu
    else :
        return []
\end{sageblock}\\
We illustrate bellow the output of the function call with the inputs
1 and 2.
\begin{equation}
\mbox{FaT}\left(1\right)=\sage{FaT(1)}.
\end{equation}
\begin{equation}
\mbox{FaT}\left(2\right)=\sage{FaT(2)}.
\end{equation}
The formulas returned by the FaT procedure are in binary tree form.
For convenience we may implement a function which output the expression
in prefix notation, the function for formatting the encoding into
prefix is provided bellow\\
\begin{sageblock}
@cached_function
def FaPre(n):
    """
    The procedure outputs the list of Formula in prefix
    notation constructed using fan-in two addition gates 
    having inputs restricted to the integral unit 1
    and the resulting formula evaluates to the input 
    integer n > 0.

    EXAMPLES:
    The input n must be greater than 0
    ::
        sage: FaPre(3)
        ['+1+11', '++111']


    AUTHORS:
    - Edinah K. Gnang, Maksym Radziwill and Doron Zeilberger

    To Do :
    - Try to implement faster version of this procedure

    """
    return [T2Pre(g) for g in FaT(n)]
\end{sageblock}\\
The postfix variant of the function implemented is immediate and provided
bellow.\\
\begin{sageblock}
@cached_function
def FaP(n):
    """
    The set of formula only using addition gates
    which evaluates to the input integer n in prefix notation.

    EXAMPLES:
    The input n must be greater than 0
    ::
        sage: FaP(3)
        ['11+1+', '111++']


    AUTHORS:
    - Edinah K. Gnang, Maksym Radziwill and Doron Zeilberger

    To Do :
    - Nothing as this procedure is optimal

    """
    return [T2P(g) for g in FaT(n)]
\end{sageblock}\\
Having implemented procedures which produces formulas using only addition,
we now turn to the problem of enumerating such formulas. Clearly we
could enumerate the sets by first producing the formulas and then
enumerating them, but this would lead to a very inefficient use of
space and time resources. Instead we compute recurrence formulas which
determines the number of formulas encoding using only additions and
with input restricted to the integral unit $1$.\\
\begin{sageblock}
@cached_function
def Ca(n):
    """
    The procedure outputs the number of Formula-binary Trees
    constructed using fan-in two addition gates and having
    inputs restricted to the integral unit 1 and the each
    of the resulting formulas each evaluate to the input integer n > 0.

    EXAMPLES:
    The input n must be greater than 0
    ::
        sage: Ca(3)
        2

    AUTHORS:
    - Edinah K. Gnang, Maksym Radziwill and Doron Zeilberger

    To Do :
    - Try to implement faster version of this procedure

    """
    if n == 1:
        return 1
    else :
        return sum([Ca(i)*Ca(n-i) for i in range(1,n)])
\end{sageblock}\\
We illustrate the usage of the functions bellow 
\begin{equation}
\mbox{Ca}\left(1\right)=\sage{Ca(1)}.
\end{equation}
\begin{equation}
\mbox{Ca}\left(2\right)=\sage{Ca(2)}.
\end{equation}
\begin{equation}
\mbox{Ca}\left(3\right)=\sage{Ca(3)}.
\end{equation}
\begin{equation}
\mbox{Ca}\left(4\right)=\sage{Ca(4)}.
\end{equation}
\begin{equation}
\mbox{Ca}\left(5\right)=\sage{Ca(5)}.
\end{equation}
furthermore we may note that 
\begin{equation}
\mbox{Ca}\left(n\right)=\sum_{i+j=n}\mbox{Ca}\left(i\right)\,\mbox{Ca}\left(j\right),\quad\mbox{Ca}\left(1\right)=1,
\end{equation}
which would suggest that for 
\[
\mbox{Ca}\left(1\right)=1,
\]
\begin{equation}
\sum_{n\ge1}\mbox{Ca}\left(n\right)x^{n}=\sum_{n\ge1}\left(\sum_{i+j=n}\mbox{Ca}\left(i\right)\,\mbox{Ca}\left(j\right)\right)x^{n},
\end{equation}
To avoid redundancy we may choose to only list formulas for which
the second term of the tree is less or equal to the integer encoded
in the left term of the tree. We provide bellow the implementation
of the procedure . \\
\begin{sageblock}
@cached_function
def LopFaT(n):
    """
    Outputs all the formula-binary trees only using addition
    such that the first term of the addition is >= the second term.

    EXAMPLES:
    The input n must be greater than 0
    ::
        sage: LopFaT(3)
        [['+', ['+', 1, 1], 1]]


    AUTHORS:
    - Edinah K. Gnang, Maksym Radziwill and Doron Zeilberger

    To Do :
    - Try to implement faster version of this procedure

    """
    if n == 0:
        return []
    elif n == 1:
        return [1]
    else :
        gu = []
        for i in range(1,1+floor(n/2)):
            gu = gu + [['+', g1, g2] for g1 in LopFaT(n-i) for g2 in LopFaT(i)]
        return gu
\end{sageblock}\\
For outputting such formulas in prefix notation we use the function
implemented bellow\\
\begin{sageblock}
def LopFaPre(n):
    """
    Outputs all the formula-binary tree 
    which evaluate to the input integer n such that the first 
    term of the addition is >= the second term in prefix notation.

    EXAMPLES:
    The input n must be greater than 0
    ::
        sage: LopFaPre(2)
        "+11"

    AUTHORS:
    - Edinah K. Gnang, Maksym Radziwill and Doron Zeilberger

    To Do :
    - Try to implement faster version of this procedure

    """
    return [T2Pre(f) for f in LopFaT(n)]
\end{sageblock}\\
For outputting such formulas in postfix notation we use the function
implemented bellow\\
\begin{sageblock}
def LopFaP(n):
    """
    Outputs all the formula-binary tree 
    which evaluate to the input integer n such that the first 
    term of the addition is >= the second term in postfix notation.

    EXAMPLES:
    The input n must be greater than 0
    ::
        sage: LopFaP(2)
        "11+"


    AUTHORS:
    - Edinah K. Gnang, Maksym Radziwill and Doron Zeilberger

    To Do :
    - Try to implement faster version of this procedure

    """
    return [T2P(f) for f in LopFaT(n)]
\end{sageblock}\\
Similarly we provide an implementation for a distinct procedure for
enumerating formulas trees for which the second term of the tree is
less or equal to the integer encoded in the left term of the tree.
\\
\begin{sageblock}
@cached_function
def LopCa(n):
    """
    Outputs the number of formula-binary trees only using addition gates
    such that the first term of the addition is >= the second term.
    EXAMPLES:
    The input n must be greater than 0
    ::
        sage: LopCa(3)
        1


    AUTHORS:
    - Edinah K. Gnang, Maksym Radziwill and Doron Zeilberger

    To Do :
    - Try to implement faster version of this procedure

    """
    if n == 1:
        return 1
    else :
        return sum([LopCa(i)*LopCa(n-i) for i in range(1,1+floor(n/2))])
\end{sageblock}\\
In many situations, there will be way more formulas then it would
be reasonable to output in a list, however for experimental purposes
it is often sufficient to generate formulas of interest uniformly
at random. Incidentally following the Wilf Methodology we implement
a function for sampling uniformly at random formula which use only
addition gates and have input restricted to the integer 1.\\
\begin{sageblock}
def RaFaT(n):
    """
    Outputs a uniformly randomly chosen formula-binary tree 
    which evaluate to the input integer n > 0.

    EXAMPLES:
    The input n must be greater than 0
    ::
        sage: RaFat(3)
        ['+', ['+', 1, 1], 1]


    AUTHORS:
    - Edinah K. Gnang, Maksym Radziwill and Doron Zeilberger

    To Do :
    - Try to implement faster version of this procedure

    """
    if n == 0:
        return []
    if n == 1:
        return [1]
    else :
        # Rolling the Loaded Die.
        j = RollLD([Ca(i)*Ca(n-i) for i in range(1,n+1)])
        return ['+', RaFaT(j), RaFaT(n-j)]
\end{sageblock}\\
Quite straightforwardly we provide bellow the implementation of the
procedure for sampling a random formulas but returning them respectively
in prefix notation \\
\begin{sageblock}
def RaFaPre(n):
    """
    Outputs a uniformly randomly chosen formula-binary tree 
    which evaluate to the input integer n in prefix notation.
    EXAMPLES:
    The input n must be greater than 0
    ::
        sage: RaFaPre(3)
        "++111"


    AUTHORS:
    - Edinah K. Gnang, Maksym Radziwill and Doron Zeilberger

    To Do :
    - Try to implement faster version of this procedure

    """
    return(T2Pre(RaFaT(n)))
\end{sageblock}\\
For outputting uniformly sampled random formula in postfix notation
we implement the function bellow\\
\begin{sageblock}
def RaFaP(n):
    """
    Outputs a uniformly randomly chosen formula-binary tree 
    which evaluate to the input integer n in postfix notation.

    EXAMPLES:
    The input n must be greater than 0
    ::
        sage: RaFaP(3)
        111++


    AUTHORS:
    - Edinah K. Gnang, Maksym Radziwill and Doron Zeilberger

    To Do :
    - Try to implement faster version of this procedure

    """
    return(T2P(RaFaT(n)))
\end{sageblock}Similarly we implement a procedure for sampling uniformly at random
a formula where the left term is greater or equal to the right term.\\
\begin{sageblock}
def RaLopFaT(n):
    """
    Outputs a uniformly randomly chosen formula-binary tree 
    which evaluate to the input integer n such that the first 
    term of the addition is >= the second term.
    EXAMPLES:
    The input n must be greater than 0
    ::
        sage: RaLopFaT(3)
        ['+', ['+', 1, 1], 1]


    AUTHORS:
    - Edinah K. Gnang, Maksym Radziwill and Doron Zeilberger

    To Do :
    - Try to implement faster version of this procedure

    """
    if n == 1:
        return [1]
    else:
        # Rolling the Loaded Die.
        j = RollLD([LopCa(i)*LopCa(n-i) for i in range(1,1+floor(n/2))])
        return ['+', RaLopFaT(n-j), RaLopFaT(j)]
\end{sageblock}For outputting a uniformly sampled formulas in prefix having it's
first term greater or equal to the second term in prefix notation
we have\\
\begin{sageblock}
def RaLopFaPre(n):
    """
    Outputs a uniformly randomly chosen formula-binary tree 
    which evaluate to the input integer n such that the first 
    term of the addition is >= the second term in Prefix notation.
    EXAMPLES:
    The input n must be greater than 0
    ::
        sage: RaLopFaPre(3)
        "++111"

    AUTHORS:
    - Edinah K. Gnang, Maksym Radziwill and Doron Zeilberger

    To Do :
    - Try to implement faster version of this procedure

    """
    return T2Pre(RaLopFaT(n))
\end{sageblock}alternatively for outputting a uniformly sampled formula with the
right term greater or equal to the left term expressed in postfix
notation we use the function implemented bellow.\\
\begin{sageblock}
def RaLopFaP(n):
    """
    Outputs a uniformly randomly chosen formula-binary tree 
    which evaluate to the input integer n such that the first 
    term of the addition is >= the second term in Postfix notation.
    EXAMPLES:
    The input n must be greater than 0
    ::
        sage: RaLopFaP(3)
        "111++"


    AUTHORS:
    - Edinah K. Gnang, Maksym Radziwill and Doron Zeilberger

    To Do :
    - Try to implement faster version of this procedure

    """
    return T2P(RaLopFaT(n))
\end{sageblock}

\subsection{Formulas only using additions and multiplications}

We discuss here in detail procedures for producing and enumerating
formulas which result from a finite combination of fan-in two addition,
multiplication gates and having inputs restricted to integer $1$.
The basic principles underlying most procedures consists in partitioning
the set of formula into disjoint sets according to the root gate of
the formulas considered. In this particular case we will consider
the partition of formulas according to wether or not the root gate
corresponds to an addition or a multiplication gate.\\
\begin{sageblock}
@cached_function
def FamTa(n):
    """
    The set of formula-binary trees only using additions and 
    multiplications gates with the root gate being an addition
    gate and most importantly evaluates to the input integer n.

    EXAMPLES:
    The input n must be greater than 0
    ::
        sage: FamTa(3)
        [['+', 1, ['+', 1, 1]], ['+', ['+', 1, 1], 1]]


    AUTHORS:
    - Edinah K. Gnang, Maksym Radziwill and Doron Zeilberger

    To Do :
    - Try to implement faster version of this procedure

    """
    if n == 0:
        return []
    elif n == 1:
        return [1]
    else :
        gu = []
        for i in range(1,n):
            gu = gu + [['+', g1, g2] for g1 in FamT(i) for g2 in FamT(n-i)]
        return gu
\end{sageblock}\\
The procedures which determines the formulas with root gate corresponding
to a multiplication gate is provided bellow :\\
\begin{sageblock}
@cached_function
def FamTm(n):
    """
    The set of formula-binary trees only using addition  and 
    multiplication gates with root gate corresponding to a 
    multiplication gate which evaluates to the input integer n.

    EXAMPLES:
    The input n must be greater than 0
    ::
        sage: FamTm(4)
        [['*',  ['+', 1, 1], ['+', 1, 1]]]


    AUTHORS:
    - Edinah K. Gnang, Maksym Radziwill and Doron Zeilberger

    To Do :
    - Try to implement faster version of this procedure

    """
    if n == 1:
        return []
    else :
        gu = []
        for i in range(2, 1+floor(n/2)):
            if mod(n,i) == 0:
                gu = gu + [['*', g1, g2] for g1 in FamT(i) for g2 in FamT(n/i)]
        return gu
\end{sageblock}\\
We implement bellow the function which compute the union of the two
partition of formulas, those rooted at an addition gate and the ones
rooted at a multiplication gate.\\
\begin{sageblock}
@cached_function
def FamT(n):
    """
    The set of formula-binary trees only using addition and 
    multiplication gates.

    EXAMPLES:
    The input n must be greater than 0
    ::
        sage: FamT(3)
        [['+', 1, ['+', 1, 1]], ['+', ['+', 1, 1], 1]]


    AUTHORS:
    - Edinah K. Gnang, Maksym Radziwill and Doron Zeilberger

    To Do :
    - Try to implement faster version of this procedure

    """
    return (FamTa(n) + FamTm(n))
\end{sageblock}\\
Again following the Wilf methodology we implement distinct procedures
for enumerating formulas which result from a finite combination of
fan-in two addition and multiplication gates. We start by implementing
the function which enumerate formulas rooted at an addition gate\\
\begin{sageblock}
@cached_function
def Cama(n):
    """
    Output the size of the set of formulas produced by the procedure FamTa(n).

    EXAMPLES:
    The input n must be greater than 0
    ::
        sage: Cama(4)
        5


    AUTHORS:
    - Edinah K. Gnang, Maksym Radziwill and Doron Zeilberger

    To Do :
    - Try to implement faster version of this procedure

    """
    if n==1:
        return 1
    else:
        return sum([Cam(i)*Cam(n-i) for i in range(1,n)])
\end{sageblock}\\
We then implement the function which enumerate formulas resulting
from finite combination of addition, multiplication gates rooted at
a multiplication gate.\\
\begin{sageblock}
@cached_function
def Camm(n):
    """
    Output the size of the set of formulas produced by the procedure FamTm(n).

    EXAMPLES:
    The input n must be greater than 0
    ::
        sage: Camm(4)
        1


    AUTHORS:
    - Edinah K. Gnang, Maksym Radziwill and Doron Zeilberger

    To Do :
    - Try to implement faster version of this procedure

    """
    if n==1:
        return 1
    else:
        return sum([Cam(i)*Cam(n/i) for i in range(2,1+floor(n/2)) if mod(n,i)==0])
\end{sageblock}\\
Finally we implement the function which enumerates all formulas which
result from a finite combination of addition, multiplication gates
which evaluate to the input integer\\
\begin{sageblock}
@cached_function
def Cam(n):
    """
    Output the size of the set of formulas produced by the procedure FamT(n).

    EXAMPLES:
    The input n must be greater than 0
    ::
        sage: Cam(6)
        52

    AUTHORS:
    - Edinah K. Gnang, Maksym Radziwill and Doron Zeilberger

    To Do :
    - Try to implement faster version of this procedure

    """
    return Cama(n)+Camm(n)
\end{sageblock}\\
As we have mentioned for formulas of large sizes we implement a function
which samples uniformly at random formulas which evaluate to the input
integer and result from a finite combination of addition and multiplication
gates and rooted at an addition gate\\
\begin{sageblock}
def RaFamTa(n):
    """
    Outputs a formula-binary tree formula sampled uniformly at random
    amoung all formulas which evaluates to the input integer n 
    the formula results from a finite combination of addition
    and multiplication gates and is rooted at an addition gate.

    EXAMPLES:
    The input n must be greater than 0
    ::
        sage: RaFamT(6)
        [['+', 1, ['+', 1, 1]], ['+', ['+', 1, 1], 1]]


    AUTHORS:
    - Edinah K. Gnang, Maksym Radziwill and Doron Zeilberger

    To Do :
    - Try to implement faster version of this procedure

    """
    if n==1:
        return 1
    else:
        j = RollLD([Cam(i)*Cam(n-i) for i in range(1,n+1)])
        return ['+', RaFamT(j), RaFamT(n-j)]
\end{sageblock}\\
Similarly we implement a function which samples a uniformly at random
a formula which evaluate to the input integer, which results from
a finite combination of addition, multiplication gates and is rooted
at a multiplication gate\\
\begin{sageblock}
def RaFamTm(n):
    """
    Outputs a formula-binary tree sampled uniformly at random
    which evaluates to the input integer n using only addition
    and multiplication gates and rooted at a mulitplication.

    EXAMPLES:
    The input n must be greater than 0
    ::
        sage: RaFamT(6)
        ['*',['+', 1, 1], ['+', ['+', 1, 1], 1]]


    AUTHORS:
    - Edinah K. Gnang, Maksym Radziwill and Doron Zeilberger

    To Do :
    - Try to implement faster version of this procedure

    """
    if n==1:
        print '1 has no multiplicative split'
        return I
    elif is_prime(n):
        print str(n)+' has no multiplicative split'
        return I
    else:
        lu = []
        L  = []
        for i in range(2,1+floor(n/2)):
            if mod(n,i)==0:
                lu.append(i)
                L.append(Cam(i)*Cam(n/i))
        j = RollLD(L)
        return ['*', RaFamT(lu[j-1]), RaFamT(n/lu[j-1])]
              
\end{sageblock}\\
Finally we can combine the two functions implemented above to obtain
a functions which samples uniformly at random a formula which evaluates
to the input integer and results from a finite combination of addition
and multiplication gate\\
\begin{sageblock}
def RaFamT(n):
    """
    Outputs a formula-binary tree sampled uniformly at random
    which evaluates to the input integer n using only addition
    and multiplication gates.

    EXAMPLES:
    The input n must be greater than 0
    ::
        sage: RaFamT(6)
        [['+', 1, ['+', 1, 1]], ['+', ['+', 1, 1], 1]]


    AUTHORS:
    - Edinah K. Gnang, Maksym Radziwill and Doron Zeilberger

    To Do :
    - Try to implement faster version of this procedure

    """
    if n==1:
        return 1
    else:
        i = RollLD[Cama(n),Camm(n)]
        if i==1:
            return RaFamTa(n)
        else :
            return RaFamTm(n)
\end{sageblock}\\
For obtaining the list all formulas which combine addition and multiplication
express using the postfix notation and evaluate to the input integer
we have \\
\begin{sageblock}
@cached_function
def FamP(n):
    """
    Outputs the set of formula-binary tree written in Postfix notation
    which evaluates to the input integer n using only addition
    and multiplication gates.

    EXAMPLES:
    The input n must be greater than 0
    ::
        sage: FamP(2)
        '11+'


    AUTHORS:
    - Edinah K. Gnang, Maksym Radziwill and Doron Zeilberger

    To Do :
    - Try to implement faster version of this procedure

    """
    return [T2P(f) for f in FamT(n)]
\end{sageblock}\\
Similarly for obtaining the list all formulas which combine addition
and multiplication gates and evaluate to the input integer express
in the prefix notation we have\\
\begin{sageblock}
@cached_function
def FamPre(n):
    """
    Outputs the set of formula-binary tree written in prefix notation
    which evaluates to the input integer n using only addition
    and multiplication gates.

    EXAMPLES:
    The input n must be greater than 0
    ::
        sage: FamPre(6)
        [['+', 1, ['+', 1, 1]], ['+', ['+', 1, 1], 1]]


    AUTHORS:
    - Edinah K. Gnang, Maksym Radziwill and Doron Zeilberger

    To Do :
    - Try to implement faster version of this procedure

    """
    return [T2Pre(f) for f in FamT(n)]
\end{sageblock}\\
For obtaining the randomly sample integer which evaluates to the input
integer and is uniformly sampled among all formulas which combine
addition and multiplication express using the postfix notation we
have \\
\begin{sageblock}
@cached_function
def RaFamP(n):
    """
    Outputs a uniformly randomly sample formula-binary tree written 
    in postfix notation which evaluates to the input integer n using 
    only addition and multiplication gates.

    EXAMPLES:
    The input n must be greater than 0
    ::
        sage: RaFamP(6)
        [['+', 1, ['+', 1, 1]], ['+', ['+', 1, 1], 1]]


    AUTHORS:
    - Edinah K. Gnang, Maksym Radziwill and Doron Zeilberger

    To Do :
    - Try to implement faster version of this procedure

    """
    return T2P(RaFamT(n))
\end{sageblock}\\
Similarly obtaining the randomly sample integer which evaluates to
the input integer and is uniformly sampled among all formulas which
combine addition and multiplication express using the prefix notation
we have \\
\begin{sageblock}
@cached_function
def RaFamPre(n):
    """
    Outputs a uniformly randomly sample formula-binary tree written 
    in prefix notation which evaluates to the input integer n using 
    only addition and multiplication gates.

    EXAMPLES:
    The input n must be greater than 0
    ::
        sage: RaFamPre(6)
        [['+', 1, ['+', 1, 1]], ['+', ['+', 1, 1], 1]]


    AUTHORS:
    - Edinah K. Gnang, Maksym Radziwill and Doron Zeilberger

    To Do :
    - Try to implement faster version of this procedure

    """
    return T2Pre(RaFamT(n))
\end{sageblock}

\subsection{Formulas only using additions, multiplications and exponentiation}

We discuss here procedures for producing and enumerating formulas
using a combination of fan-in two addition, multiplication and exponentiation
gates. the principles used are very much analogous to those used in
the previous section. We start by formulas rooted at addition gates
\\
\begin{sageblock}
@cached_function
def FameTa(n):
    """
    The set of formula-binary trees only using addition,
    multiplication, and exponentiation gates. The root gate
    being an addition gate and and the formula evaluates to
    the input integer n.

    EXAMPLES:
    The input n must be greater than 0
    ::
        sage: FameTa(2)
        ['+', 1, 1]


    AUTHORS:
    - Edinah K. Gnang, Maksym Radziwill and Doron Zeilberger

    To Do :
    - Try to implement faster version of this procedure

    """
    if n == 1:
        return [1]
    else:
        gu = []
        for i in range(1,n):
            gu = gu + [['+', g1, g2] for g1 in FameT(i) for g2 in FameT(n-i)]
        return gu
\end{sageblock}\\
next we implement procedure for listing formulas rooted at a multiplication
gate\\
\begin{sageblock}
@cached_function
def FameTm(n):
    """
    The set of formula-binary trees only using addition. 
    multiplication and exponentiation gates with the top 
    gate being a multiplication gate which evaluates to the 
    input integer n.

    EXAMPLES:
    The input n must be greater than 0
    ::
        sage: FameTm(3)
        [['+', 1, ['+', 1, 1]], ['+', ['+', 1, 1], 1]]


    AUTHORS:
    - Edinah K. Gnang, Maksym Radziwill and Doron Zeilberger

    To Do :
    - Try to implement faster version of this procedure

    """
    if n == 1:
        return []
    else :
        gu = []
        for i in range(2,1+floor(n/2)):
            if mod(n,i) == 0:
                gu = gu + [['*', g1, g2] for g1 in FameT(i) for g2 in FameT(n/i)]
        return gu
\end{sageblock}\\
and finally we list formulas rooted at an exponentiation gates\\
\begin{sageblock}
@cached_function
def FameTe(n):
    """
    The set of formula-binary trees only using addition. 
    multiplication and exponentiation gates with the top 
    gate being an exponetiation gate which evaluates to the 
    input integer n.

    EXAMPLES:
    The input n must be greater than 0
    ::
        sage: FameTe(3)
        [['+', 1, ['+', 1, 1]], ['+', ['+', 1, 1], 1]]


    AUTHORS:
    - Edinah K. Gnang, Maksym Radziwill and Doron Zeilberger

    To Do :
    - Try to implement faster version of this procedure

    """
    if n == 1:
        return []
    else :
        gu = []
        for i in range(2,2+floor(log(n)/log(2))):
            if floor(n^(1/i)) == ceil(n^(1/i)):
                gu = gu + [['^', g1, g2] for g1 in FameT(i) for g2 in FameT(n^(1/i))]
        return gu
\end{sageblock}\\
Finally combining the three function implemented above we obtain the
function which lists all formulas which combine addition, multiplication,
and exponentiation gates which evaluate to the input integer.\\
\begin{sageblock}
@cached_function
def FameT(n):
    """
    The set of formula-binary trees only using addition. 
    multiplication and exponentiation gates which evaluates to the 
    input integer n.

    EXAMPLES:
    The input n must be greater than 0
    ::
        sage: FameT(3)
        [['+', 1, ['+', 1, 1]], ['+', ['+', 1, 1], 1]]


    AUTHORS:
    - Edinah K. Gnang, Maksym Radziwill and Doron Zeilberger

    To Do :
    - Try to implement faster version of this procedure

    """
    return FameTa(n) + FameTm(n) + FameTe(n)
\end{sageblock}\\
For a more efficient enumeration of the formulas resulting from combination
of addition, multiplication and exponentitation gates which evaluate
to the input integer we consider here enumerating procedure for formulas
rooted at the addition gate:\\
\begin{sageblock}
@cached_function
def Camea(n):
    """
    Output the size of the set of formulas produced by the procedure FamTa(n).

    EXAMPLES:
    The input n must be greater than 0
    ::
        sage: Camea(6)
        [['+', 1, ['+', 1, 1]], ['+', ['+', 1, 1], 1]]


    AUTHORS:
    - Edinah K. Gnang, Maksym Radziwill and Doron Zeilberger

    To Do :
    - Try to implement faster version of this procedure

    """
    if n==1:
        return 1
    else:
        return sum([Came(i)*Came(n-i) for i in range(1,n)])
\end{sageblock}\\
then rooted at a multiplication gate \\
\begin{sageblock}
@cached_function
def Camem(n):
    """
    Output the size of the set of formulas produced by the procedure FamTa(n).

    EXAMPLES:
    The input n must be greater than 0
    ::
        sage: Camm(6)
        [['+', 1, ['+', 1, 1]], ['+', ['+', 1, 1], 1]]


    AUTHORS:
    - Edinah K. Gnang, Maksym Radziwill and Doron Zeilberger

    To Do :
    - Try to implement faster version of this procedure

    """
    if n==1:
        return 1
    else:
        return sum([Came(i)*Came(n/i) for i in range(2,1+floor(n/2)) if mod(n,i)==0])
\end{sageblock}\\
then rooted at an exponentiation gate \\
\begin{sageblock}
@cached_function
def Camee(n):
    """
    Output the size of the set of formulas produced by the procedure FamTa(n).

    EXAMPLES:
    The input n must be greater than 0
    ::
        sage: Camee(6)
        [['+', 1, ['+', 1, 1]], ['+', ['+', 1, 1], 1]]


    AUTHORS:
    - Edinah K. Gnang, Maksym Radziwill and Doron Zeilberger

    To Do :
    - Try to implement faster version of this procedure

    """
    if n==1:
        return 1
    else:
        return sum([Came(i)*Came(n^(1/i)) for i in range(2,2+floor(log(n)/log(2)))\
if floor(n^(1/i)) == ceil(n^(1/i))])
\end{sageblock}The enumeration scheme can be described using non-linear recurrence
formula expressed earlier and repeated here for the convenience of
the reader
\begin{equation}
C_{\left\{ +,\times,\mathcircumflex\right\} }^{\left\{ +\right\} }\left(n\right)=\sum_{0<i<n}C_{\left\{ +,\times,\mathcircumflex\right\} }^{\left\{ +,\times,\mathcircumflex\right\} }\left(i\right)\: C_{\left\{ +,\times,\mathcircumflex\right\} }^{\left\{ +,\times,\mathcircumflex\right\} }\left(n-i\right)
\end{equation}
\begin{equation}
C_{\left\{ +,\times,\mathcircumflex\right\} }^{\left\{ \times\right\} }\left(n\right)=\sum_{\begin{array}{c}
1<i<\left\lfloor \frac{n}{2}\right\rfloor \\
n\,|\, i
\end{array}}C_{\left\{ +,\times,\mathcircumflex\right\} }^{\left\{ +,\times,\mathcircumflex\right\} }\left(i\right)\: C_{\left\{ +,\times,\mathcircumflex\right\} }^{\left\{ +,\times,\mathcircumflex\right\} }\left(i^{(-1)}n\right)
\end{equation}
\begin{equation}
C_{\left\{ +,\times,\mathcircumflex\right\} }^{\left\{ \mathcircumflex\right\} }\left(n\right)=\sum_{\begin{array}{c}
1<i<\left\lfloor \frac{n}{2}\right\rfloor \\
\left\lfloor n^{i^{(-1)}}\right\rfloor =\left\lceil n^{i^{(-1)}}\right\rceil 
\end{array}}C_{\left\{ +,\times,\mathcircumflex\right\} }^{\left\{ +,\times,\mathcircumflex\right\} }\left(i\right)\: C_{\left\{ +,\times,\mathcircumflex\right\} }^{\left\{ +,\times,\mathcircumflex\right\} }\left(n^{i^{(-1)}}\right)
\end{equation}
and 
\begin{equation}
C_{\left\{ +,\times,\mathcircumflex\right\} }^{\left\{ +,\times,\mathcircumflex\right\} }\left(n\right)=\sum_{\mathfrak{g}\in\left\{ +,\times,\mathcircumflex\right\} }C_{\left\{ +,\times,\mathcircumflex\right\} }^{\left\{ \mathfrak{g}\right\} }\left(n\right)
\end{equation}
so that procedure which enumerate formulas evaluating to the input
integer and resulting from finite combination of addition, multiplication
and exponentitation gates is implemented bellow\\
\begin{sageblock}
@cached_function
def Came(n):
    """
    Output the size of the set of formulas produced by the procedure FamTa(n).

    EXAMPLES:
    The input n must be greater than 0
    ::
        sage: Came(6)
        [['+', 1, ['+', 1, 1]], ['+', ['+', 1, 1], 1]]


    AUTHORS:
    - Edinah K. Gnang, Maksym Radziwill and Doron Zeilberger

    To Do :
    - Try to implement faster version of this procedure

    """
    return Camea(n)+Camem(n)+Camee(n)
\end{sageblock}\\
The code for computing the base of the exponent in the asymptotic
formula, when exponentiation gates are not allowed\\
\begin{sageblock}
def ConstanI(nb_terms, nb_itrs, prec):
	# expressing the truncated series
    f = sum([Cam(n)*x^n for n in range(1,nb_terms)])
    g = sum([Cam(d)*(f.subs(x=(x^d))-x^d) for d in range(2,nb_terms)])
    g = 1/4-g
    xk = 1/4.077
    for itr in range(nb_itrs):
        xkp1 = RealField(prec)(g.subs(x=xk))
        xk   = xkp1
    return RealField(prec)(1/xk)
\end{sageblock}\\
The code for computing the base of the exponent in the asymptotic
formula, when exponentiation gates are allowed\\
\begin{sageblock}
def ConstanII(nb_terms, nb_itrs, prec):
	# expressing the truncated series
    f = sum([Came(n)*x^n for n in range(1,nb_terms)])
    g = sum([Came(d)*(f.subs(x=(x^d))-x^d) for d in range(2,nb_terms)])
    g = 1/4-g
    xk = 1/4.131
    for itr in range(nb_itrs):
        xkp1 = RealField(prec)(g.subs(x=xk))
        xk   = xkp1
    return RealField(prec)(1/xk)
\end{sageblock}\\
Code for computing the constant factor multiple in the asymptotic
formula\\
\begin{sageblock}
def ConstanIII(nb_terms, nb_itrs, prec):
    f = sum([Cam(n)*x^n for n in range(1,100)])
    g = sum([Cam(d)*(f.subs(x=(x^d))-x^d) for d in range(2,100)])
    g1 = 1/4-g
    # Iteration
    xk = 1/4.077
    for itr in range(20):
        xkp1 = RealField(100)(g1.subs(x=xk))
        xk   = xkp1
        print RealField(100)(1/xk)
    # Setting the constant rho
    r = xk
    h = x + g
    G = expand((1-4*h)*sum([(x/r)^j for j in range(100)]))
    L = G.operands()
    Ls = []
    for i in range(100):
        Ls.append(L[len(L)-i-1])
    G = sum(Ls)
    G1 = sqrt(G.subs(x = x*r))
    c  =-1/2/sqrt(pi)
    print N(-G1.subs(x=1)*c/2)
    C = N(-G1.subs(x=1)*c/2)
    # Computing the list of ratio for ploting.
    Rt = [Cam(n)*sqrt(n^3)/(C*(1/r)^n) for n in range(2,100)]
    Plt = line([(n,N(Rt[n])) for n in range(len(Rt))])
    return [Plt,Rt]
\end{sageblock}

\section{Shortest Formulas}

Finally we use dynamic programming to determine the shortest monotone
formula which evaluates to input integers. \\
\begin{sageblock}
@cached_function
def ShortestTame(n):
    """
    Outputs the length and an example of the smallest binary-tree 
    formula using fan-in two addition, multiplication and 
    exponentiation gates.

    EXAMPLES:
    The input n must be greater than 0
    ::
        sage: ShortestTame(6)
        [9, ['*', ['+', 1, 1], ['+', 1, ['+', 1, 1]]]]


    AUTHORS:
    - Edinah K. Gnang, Maksym Radziwill and Doron Zeilberger

    To Do :
    - Try to implement faster version of this procedure

    """
    if n==1:
        return [1,1]
    else:
        aluf = []
        si = 2*n
        for i in range(1,n):
            T1 = ShortestTame(i)
            T2 = ShortestTame(n-i)
            if (T1[0]+T2[0]+1) < si:
                si = T1[0]+T2[0]+1
                if Eval(T1[1]) <= Eval(T2[1]):
                    aluf = ['+', T1[1], T2[1]]
                else:
                    aluf = ['+', T2[1], T1[1]]

        for i in range(2,floor(n/2)):
            if mod(n,i)==0:
                T1 = ShortestTame(i)
                T2 = ShortestTame(n/i)
                if (T1[0]+T2[0]+1) < si:
                    si = T1[0]+T2[0]+1
                    if Eval(T1[1]) <= Eval(T2[1]):
                        aluf = ['*', T1[1], T2[1]]
                    else:
                        aluf = ['*', T2[1], T1[1]]

        for i in range(2,2+floor(log(n)/log(2))):
            if floor(n^(1/i)) == ceil(n^(1/i)):
                T1 = ShortestTame(n^(1/i))
                T2 = ShortestTame(i)
                if (T1[0]+T2[0]+1) < si:
                    si = T1[0]+T2[0]+1
                    aluf = ['^', T1[1], T2[1]]
        return [si, aluf]
\end{sageblock}The recurrence formula scheme for determining the minimal formula
encoding is given by the following tropicalization of the enumeration
recurrence formula
\begin{equation}
\begin{cases}
\begin{array}{ccc}
S^{+}\left(n\right) & = & \min_{k}\left\{ 1+S\left(k\right)+S\left(n-k\right)\right\} \\
S^{\times}\left(n\right) & = & \min_{k}\left\{ 1+S\left(k\right)+S\left(n\cdot k^{-1}\right)\right\} \\
S^{\mathcircumflex}\left(n\right) & = & \min_{k}\left\{ 1+S\left(n\right)+S\left(n^{\left(k^{-1}\right)}\right)\right\} 
\end{array}\end{cases}
\end{equation}
and 
\begin{equation}
S\left(n\right)=\min\left\{ S^{+}\left(n\right),\: S^{\times}\left(n\right),\: S^{\mathcircumflex}\left(n\right)\right\} 
\end{equation}

\section{Goodstein encodings}

Throughout the discussion the special formula $1+1$ occurs often
enough to deserve an abbreviation, we shall use here the symbol $x$,
incidentally it is immediate that the our formula encoding can be
viewed as functions and this fact will of some significance in subsequent
discussion. But first as we have introduced our canonical encodings
let us describe two natural algorithms for recovering formula encoding
for relatively large set of integers. For computing Goodstein canonical
forms for relatively large set of integers we consider the following
set recurrence defined by
\begin{equation}
N_{0}=\left\{ 1\right\} 
\end{equation}
\begin{equation}
N_{t+1}=\bigcup_{S\subset\left\{ \left\{ 0\right\} \cup N_{t}\right\} \backslash\left\{ \emptyset\right\} }\sum_{s\in S}x^{s}
\end{equation}
note that for $k>1$, we have 
\begin{equation}
\left|N_{k}\right|=\quad^{k}2.
\end{equation}
the implementation of the recurrence is just as straight forward.\\
\begin{sageblock}
def goodstein(number_of_iterations=1):
    """
    Produces the set of symbolic expressions associated with the
    the first canonical form. In all the expressions the symbolic
    variable x stands for a short hand notation for the formula (1+1). 

    ::
        sage: goodstein(1)
        [1, x^x, x, x^x + 1, x + 1, x^x + x, x^x + x + 1]

    AUTHORS:
    - Edinah K. Gnang, Maksym Radziwill and Doron Zeilberger

    To Do :
    - Try to implement faster version of this procedure

    """
    # Initial condition of Initial set
    N0 = [1, x]

    # Main loop performing the iteration
    for iteration in range(number_of_iterations):
		# Implementation of the set recurrence
        N0 = [1] + [x^n for n in N0]
		# Initialization of a buffer list N1
        # which will store updates to N0
        N1 = []
        for n in Set(N0).subsets():
            if n.cardinality() > 0:
                N1.append(sum(n))
        N0 = list(N1)
    return N0
\end{sageblock}\\
As illustration for the computation 
\begin{equation}
N_{1}=\sage{goodstein(1)}
\end{equation}
One of the major benefit of the Goodstein encoding is the fact the
additional transformation rule 
\begin{equation}
\begin{array}{c}
1+1\leftrightarrows x\end{array}
\end{equation}
results in the classical algorithms for integer addition, multiplication
and exponentiation. In other words the Goodstein encoding unifies
into a single algorithm the seemingly different decimal algorithms
for addition, multiplication and exponentiation, the price we pay
for such a convenience is a factor $O\left(\log\log\left(n\right)\right)$
additional space for encoding the integers.

\subsubsection*{Example:}

Let us illustrate the general principle by recovering the Goodstein
encoding for the number encoded by the formula $\left(x^{x}+1\right)^{\left(x+1\right)}$
the main steps of the sequence of transformations are thus sketch
bellow: \\
$\left(x^{x}+1\right)\left(x^{x}+1\right)^{1+1}\longrightarrow\left(x^{x}x^{x}+x^{x}+x^{x}+1\right)\left(x^{x}+1\right)\longrightarrow\left(x^{x^{x}}+x^{x+1}+1\right)\left(x^{x}+1\right)\longrightarrow x^{x^{x}+x}+x^{x^{x}+1}+x^{x^{x}}+x^{x+1}+x+1$

\section{Zeta recursion and the combinatorial tower sieve }

Second Canonical Form (SCF) encoding are derived from the zeta recursion.
\begin{equation}
\check{\mathbb{N}}_{1}\::=\left\{ 1\right\} \cup\mathbb{P}_{1}\,:=\left\{ 2\right\} 
\end{equation}
\begin{equation}
\mathbb{N}_{k+1}^{(0)}=\check{\mathbb{N}}_{k}\cup\left(\left]^{k}2,\:2^{\left(^{\left(k-1\right)}2+1\right)}\right]\cap\prod_{p\in\mathbb{P}_{k}}\left\{ 1\cup p^{\check{\mathbb{N}}_{k}\cap\left[1,\log_{p}\left\{ 2^{\left(^{\left(k-1\right)}2+1\right)}\right\} \right]}\right\} \right)
\end{equation}
and $\check{\mathbb{N}}_{k+1}^{(0)}$ is deduced from $\mathbb{N}_{k+1}^{(0)}$
via completion and hence 
\begin{equation}
\mathbb{P}_{k+1}^{(0)}=\mathbb{P}_{k}\cup\left(\check{\mathbb{N}}_{k+1}^{(0)}\backslash\mathbb{N}_{k+1}^{(0)}\right)
\end{equation}
more generally we have that 
\begin{equation}
\forall\:0\le t<\left(^{k}2-{}^{\left(k-1\right)}2\right),\quad\mathbb{N}_{k+1}^{(t+1)}=\check{\mathbb{N}}_{k+1}^{(t)}\cup\left(\left]2^{\left(^{k-1}2+t\right)},\:2^{\left(^{k-1}2+t+1\right)}\right]\cap\prod_{p\in\mathbb{P}_{k+1}^{(t)}}\left\{ \left\{ 1\right\} \cup p^{\check{\mathbb{N}}_{k}\cap\left[1,\log_{p}\left\{ 2^{\left(^{k-1}2+t+1\right)}\right\} \right]}\right\} \right)
\end{equation}
quite similarly $\check{\mathbb{N}}_{k+1}^{(t+1)}$ is deduced from
$\mathbb{N}_{k+1}^{(t+1)}$ via completion and hence 
\begin{equation}
\mathbb{P}_{k+1}^{(t+1)}=\mathbb{P}_{k+1}^{(t)}\cup\left(\check{\mathbb{N}}_{k+1}^{(t+1)}\backslash\mathbb{N}_{k+1}^{(t+1)}\right),
\end{equation}
finally 
\begin{equation}
\check{\mathbb{N}}_{k+1}\::=\check{\mathbb{N}}_{k+1}^{({}^{k}2-{}^{\left(k-1\right)}2)},\mbox{ and}\quad\mathbb{P}_{k+1}\::=\mathbb{P}_{k+1}^{({}^{k}2-{}^{\left(k-1\right)}2)}
\end{equation}
The associated rational subset construction $\mathbb{Q}_{k}$ is specified
by 
\begin{equation}
\mathbb{Q}_{k+1}=\prod_{p\in\mathbb{P}_{k+1}}\left\{ \left(p^{-1}\right)^{\check{\mathbb{N}}_{k+1}}\cup\left\{ 1\right\} \cup p^{\check{\mathbb{N}}_{k+1}}\right\} .
\end{equation}
The implementation of the zeta recurrence is therefore given by\\
\begin{sageblock}
def SCF(nbitr):
    # Symbol associated with the prime 2.
    x = var('x')
    # Pr corresponds to the initial list of primes
    Pr = [x]
    # Nu corresponds to the initial list of integer
    NuC  = [1,x]; TNuC = [1,x]
    # Initializing the upper and lower bound
    upr_bnd = 2^2; lwr_bnd = 2
    # Computing the set recurrence
    for itr in range(nbitr):
        for jtr in range(log(upr_bnd,2)-log(lwr_bnd,2)):
            TpNu = [1]
            for p in Pr:
                 TpNu=TpNu+\
[m*pn for m in TpNu for pn in [p^n for n in NuC if (p^n).subs(x=2)<=\
2^(N(log(lwr_bnd,2))+jtr+1)] if (m*pn).subs(x=2)<=2^(N(log(lwr_bnd,2))+jtr+1)]
            # Keeping the elements within the range of the upper and lower bound
            Nu = [f for f in TpNu if (2^(N(log(lwr_bnd,2))+jtr)<\
f.subs(x=2) and f.subs(x=2)<=2^(N(log(lwr_bnd,2))+jtr+1))]
            print '\nThe iteration will find '+\
str(2^(N(log(lwr_bnd,2))+jtr+1)-2^(N(log(lwr_bnd,2))+jtr)-len(Nu))+\
' new primes in ['+str(2^(N(log(lwr_bnd,2))+jtr))+\
', '+str(2^(N(log(lwr_bnd,2))+jtr+1))+']'
            # Obtaining the corresponding sorted integer list
            la = [f.subs(x=2) for f in Nu]; lb = copy(la); lb.sort()
            # Obtaining the sorting permutation
            perm = []
            for i1 in range(len(la)):
                for i2 in range(len(lb)):
                    if lb[i1]==la[i2]:
                        perm.append(i2)
                        break
            # Sorting the list using the obtained permutation
            Nu = [Nu[perm[j]] for j in range(len(Nu))]
            # Computing the set completion
            TNuC = TNuC + Nu
            l = len(TNuC)
            i = 2^(log(lwr_bnd,2)+jtr)-1
            while i<l-1:
                if(TNuC[i+1].subs(x=2)-TNuC[i].subs(x=2)==2):
                    Pr.append(TNuC[i]+1)
                    TNuC.insert(i+1,TNuC[i]+1)
                    l=l+1
                else:
                    i=i+1
        # Updating the list of integers
        NuC = TNuC
        # Updating the upper and lower bound
        lwr_bnd = upr_bnd; upr_bnd = 2^upr_bnd
    return [Pr, NuC]                    
\end{sageblock}\\
We deduce from the Similarly the code for obtaining SCF encodings
for rational numbers is provided bellow. \\
\begin{sageblock}
def RationalSet(Pr, NuC):
    # Initialization of the rational set
    QuC = [1]
    # Computing the set
    for p in Pr:
        QuC=QuC+[m*pn for m in QuC for pn in [p^n for n in NuC]+\
[p^(-n) for n in NuC]]
    return QuC
\end{sageblock}\\
If our in main interest is however to sieve out only SCF encodings
of primes, we would consider the following slightly modified zeta
recursion 
\begin{equation}
\mathbb{N}_{2,k+1}=\bigcup_{\begin{array}{c}
n\in\mathbb{\check{N}}_{k}\\
2^{k+1}<2^{n}\le2^{k+2}
\end{array}}2^{n}.
\end{equation}
$\forall q\in\mathbb{P}_{k}$ such that $q>2$ we consider the sets
\begin{equation}
\mathbb{N}_{1,q,k+1}=\bigcup_{\begin{array}{c}
n\in\mathbb{\check{N}}_{k}\\
2<q^{n}\le2^{k+2}
\end{array}}q^{n}
\end{equation}
\begin{equation}
\mathbb{N}_{2,q,k+1}=\bigcup_{\begin{array}{c}
n\in\mathbb{N}_{1,q,k+1}\\
n<np^{m}<2^{k+2}\\
p\in\mathbb{P}_{k},\mbox{ and }p<q
\end{array}}n\, p^{m}
\end{equation}
\[
\vdots
\]
\begin{equation}
\mathbb{N}_{t+1,q,k+1}=\bigcup_{\begin{array}{c}
n\in\mathbb{N}_{t,q,k+1}\\
n<np^{m}<2^{k+2}\\
p\in\mathbb{P}_{k},\mbox{ and }p<q
\end{array}}np^{m}
\end{equation}
\[
\vdots
\]
\begin{equation}
\mathbb{N}_{\left|\mathbb{P}_{k}\right|,q,k+1}=\bigcup_{\begin{array}{c}
n\in\mathbb{N}_{\left|\mathbb{P}_{k}\right|-1,\, q,k+1}\\
2^{k+1}<np^{m}<2^{k+2}
\end{array}}n\, p^{m}
\end{equation}
and hence 
\begin{equation}
\forall q\in\mathbb{P}_{k}\backslash\left\{ 2\right\} ,\quad\mathbb{N}_{q,\, k+1}=\bigcup_{0<i\le\pi\left(q\right)}\mathbb{N}_{i,q,k+1}
\end{equation}
furthermore we have 
\begin{equation}
\left[2^{k+1},\,2^{k+2}\right]\cap\mathbb{N}_{k+1}=\bigcup_{q\in\mathbb{P}_{k}}\mathbb{N}_{q,k+1}
\end{equation}
Finally, the set completion of $\mathbb{N}_{k+1}$ to $\check{\mathbb{N}}_{k+1}$
is obtained by adjoining to the set $\mathbb{N}_{k+1}$ formula integer
encodings of the form $1+\min\left\{ m,n\right\} $, for all unordered
pairs $\left(m,n\right)$ of distinct elements of $\mathbb{N}_{k+1}$
such that 
\begin{equation}
\nexists\; j\in\mathbb{N}_{k+1}\mbox{ with }\;\min\left\{ m,n\right\} <j<\max\left\{ m,n\right\} =2+\min\left\{ m,n\right\} .
\end{equation}
The implementation of the modified zeta recursion as discussed above
is discussed bellow\\
\begin{sageblock}
def N_1_k_plus_1(Nk, Pk, k):
    L = []
    for q in Pk:
        for n in range(floor(ln(2^(k+1))/ln(q.subs(x=2))), floor(ln(2^(k+2))/ln(q.subs(x=2)))):
            L.append(q^Nk[n])
    return L
\end{sageblock}then we consider procedure bellow which generates a script for constructing
composite tower with a given number of factors\\
\begin{sageblock}
def generate_factor_script(c):
    # Creating the string corresponding to the file name
    filename = 'N_'+str(c)+'_kplus1.sage'
   # Opening the file
    f = open(filename,'w')
    f.write('def N_'+str(c)+'_k_plus_1(Nk, Pk, k):\n')
    f.write('    L = []\n')
    # variable storing the spaces
    sp = ''
    for i in range(c):
        if i<1:
            sp=sp+'    '
            f.write(sp+'for p'+str(i)+' in Pk:\n')
            sp=sp+'    '
            f.write(sp+'for n'+str(i)+' in range(floor(ln(2^(k+2))/ln(p'+str(i)+'.subs(x=2)))):\n' )
        elif i==c-1:
            sp=sp+'    '
            f.write(sp+'for p'+str(i)+' in Pk[Pk.index(p'+str(i-1)+')+1:]:\n')
            sp=sp+'    '
            dv = ''
            for d in range(i):
                # string keeping track of the divisors
                if d == i-1:
                    dv=dv+'(p'+str(i-1)+'^Nk[n'+str(i-1)+']).subs(x=2)'
                else:
                    dv=dv+'(p'+str(d)+'^Nk[n'+str(d)+']).subs(x=2)*'
            f.write(sp+'if floor(ln(2^(k+1)/('+dv+'))/ln(p'+str(i)+'.subs(x=2)))>=0:\n')
            sp=sp+'    '
            f.write(sp+'for n'+str(i)+' in range(floor(ln(2^(k+1)/('+dv+'))/ln(p'+str(i)+'.subs(x=2))),\
floor(ln(2^(k+2)/('+dv+'))/ln(p'+str(i)+'.subs(x=2)))):\n')
            sp=sp+'    '
            mt = ''
            for d in range(c):
                # string keeping track of the symbolic SCF expression
                if d == c-1:
                    mt=mt+'p'+str(c-1)+'^Nk[n'+str(c-1)+']'
                else:
                    mt=mt+'p'+str(d)+'^Nk[n'+str(d)+']*'
            f.write(sp+'L.append('+mt+')\n    return L')
        else:
            sp=sp+'    '
            f.write(sp+'for p'+str(i)+' in Pk[Pk.index(p'+str(i-1)+')+1:]:\n')
            sp=sp+'    '
            dv = ''
            for d in range(i):
                # string keeping track of the divisors
                if d==i-1:
                    dv=dv+'(p'+str(i-1)+'^Nk[n'+str(i-1)+']).subs(x=2)'
                else:
                    dv=dv+'(p'+str(d)+'^Nk[n'+str(d)+']).subs(x=2)*'
            f.write(\
sp+'for n'+str(i)+' in range(floor(ln(2^(k+2)/('+dv+'))/ln(p'+str(i)+'.subs(x=2)))):\n')
    # Closing the file
    f.close()	
\end{sageblock}then the main procedure which uses the two procedure implemented above
is implemented here \\
\begin{sageblock}
def zetarecursionII(nbitr):
    # Defining the symbolic variables x which corresponds
    # to shorthand notation for (1+1).
    var('x') 
    # Initial conditions for the zeta recursion.
    # Initial list of primes in SCF encoding
    Pi = [x]
    # Initial list of expression associated with the SCF
    # integer encoding.
    Ni = [1] + Pi
    if nbitr == 0:
        return [Ni, Pi, i]
    # The first iteration properly starts here
    i = 0
    Rb = []
    Rb.append(Ni[len(Ni)-1])
    Rb = Rb + N_1_k_plus_1(Ni, Pi, i)
    # Sorting the obtainted list
    Tmp = []
    for f in range(2^(i+1),2^(i+2)+1):
        Tmp.append([])
    for f in Rb:
        Tmp[-2^(i+1)+f.subs(x=2)].append(f)
    # Filling up Rb in order
    Rb = []
    for f in range(len(Tmp)):
        if len(Tmp[f]) == 1:
            Rb.append(Tmp[f][0])
        else:
            Rb.append(Tmp[f-1][0]+1)
            Pi.append(Tmp[f-1][0]+1)
    Ni = list(Ni+Rb[1:])
    if nbitr == 1:
        return [Ni, Pi, i]
    for i in range(1, nbitr+1):
        print 'Iteration number '+str(i)
        Rb = []
        Rb.append(Ni[len(Ni)-1])
        Rb = Rb + N_1_k_plus_1(Ni, Pi, i)
        # Code for going beyound a single prime factors 
        prm = 6
        c = 2
        while  prm < 2^(i+2):
            generate_factor_script(c)
            load('N_'+str(c)+'_kplus1.sage')
            Rb = Rb + eval("N_%d_k_plus_1(Ni,Pi,%d)"%(c,i))
            # Since ironically c indexes the next prime we have 
            prm = prm*Integer((Pi[c-1]).subs(x=2))
            c = c+1
        # Sorting the obtainted list
        Tmp = []
        for f in range(2^(i+1),2^(i+2)+1):
            Tmp.append([])
        for f in Rb:
            Tmp[-2^(i+1)+f.subs(x=2)].append(f)
        # Filling up Rb in order
        Rb = []
        for f in range(len(Tmp)):
            if len(Tmp[f]) == 1:
                Rb.append(Tmp[f][0])
            else:
                Rb.append(Tmp[f-1][0]+1)
                Pi.append(Tmp[f-1][0]+1)
        Ni = list(Ni+Rb[1:])
    return [Ni, Pi, i]
\end{sageblock}then running the procedure yields the following set of primes\\
\begin{sageblock} Lp3 = zetarecursionII(3)[1]\end{sageblock}

\[
\mathbb{P}_{3}=\left[\sage{Lp3[0]},\sage{Lp3[1]},\sage{Lp3[2]},\sage{Lp3[3]},\sage{Lp3[4]},\right.
\]
\[
\sage{Lp3[5]},\sage{Lp3[6]},\sage{Lp3[7]},\sage{Lp3[8]},\sage{Lp3[9]},
\]
\begin{equation}
\left.\sage{Lp3[8]},\sage{Lp3[9]}\right]
\end{equation}
Incidentally the number of composites less than $2^{k+2}$ with the
prime $q$ in their tower connected to the root is given by 
\begin{equation}
\sum_{q\in\mathbb{P}_{k}}\left|\mathbb{N}_{q,k+1}\right|
\end{equation}
so that we have 
\begin{equation}
\pi\left(2^{k+2}\right)-\pi\left(2^{k+1}\right)=2^{k+2}-\sum_{q\in\mathbb{P}_{k}}\left|\mathbb{N}_{q,k+1}\right|
\end{equation}

\section{Horner encoding.}

The encoding that we discuss appears to be just as natural as the
Goodstein encoding and offers the benefit of yield considerably smaller
monotone formula encodings of integers. The recursive Horner encoding
also has the advantage that that it can be efficiently deduced from
the Goodstein endcoding, this is of course not true of the SCF.\\
\begin{sageblock}
def RecursiveHorner(nbitr=1):
    x = var('x')
    Nk  = [1, x, 1+x, x^x]
    # Initialization of the lists
    LEk = [x^x]
    LOk = [1+x]
    LPk = [x, x^x]   
    # Main loop computing the encoding
    for i in range(nbitr):
        # Updating the list
        LEkp1 = [m*n for m in LPk for n in LOk] + [x^m for m in LEk+LOk]
        LOkp1 = [n+1 for n in LEk]
        LPkp1 = LPk + [x^m for m in LEk+LOk]
        # The New replaces the old
        Nk = Nk + LEkp1+LOkp1
        LEk = LEkp1
        LOk = LOkp1
        LPk = LPkp1
    return Nk
\end{sageblock}

\section*{Acknowledgments}

This material is based upon work supported by the National Science
Foundation under agreements Princeton University Prime Award No. CCF-0832797
and Sub-contract No. 00001583. The author would like to thank the
IAS for providing excellent working conditions. The author is also
grateful to Maksym Radziwill for providing the code for the computation
of the constants in the asymptotic formula, to Doron Zeilberger whos's
initial maple implementation inspired the current implementation and
to Carlo Sanna for insightful comments and suggestions while preparing
this package.

\end{document}